\documentclass[aps,prd,onecolumn,amsmath,nofootinbib,nobibnotes,superscriptaddress]{revtex4}
\def\gsim{ \lower .75ex \hbox{$\sim$} \llap{\raise .27ex \hbox{$>$}} }
\usepackage{graphicx,}

\begin{document}

\title{Detecting dark energy in long baseline neutrino oscillations}

\author{Pei-Hong Gu}
\email{guph@mail.ihep.ac.cn} \affiliation{Theoretical Physics
Division, IHEP, Chinese Academy of Sciences, Beijing 100049, P. R.
China}

\author{Xiao-Jun Bi}
\email{bixj@mail.ihep.ac.cn} \affiliation{Key laboratory of
particle astrophysics, IHEP, Chinese Academy of Sciences, Beijing
100049, P. R. China}

\author{Bo Feng}
\email{fengbo@resceu.s.u-tokyo.ac.jp} \affiliation{ National
Astronomical Observatories, Chinese Academy of Sciences, Beijing
100012, P. R. China} \affiliation{ Research Center for the Early
Universe(RESCEU), Graduate School of Science, The University of
Tokyo, Tokyo 113-0033, Japan}

\author{Bing-Lin Young}
\email{young@iastate.edu} \affiliation{Department of Physics and
Astronomy, Iowa State University, Ames, Iowa 50011, USA}

\author{Xinmin Zhang}
\email{xmzhang@mail.ihep.ac.cn} \affiliation{Theoretical Physics
Division, IHEP, Chinese Academy of Sciences, Beijing 100049, P. R.
China}

\begin{abstract}
\vskip 3ex
\begin{center} {\bf Abstract} \end{center}
In this paper, we discuss a possibility of studying properties of
dark energy in long baseline neutrino oscillation experiments. We
consider two types of models of neutrino dark energy. For one type
of models the scalar field is taken to be quintessence-like and
for the other phantom-like. In these models the scalar fields
couple to the neutrinos to give rise to a spatially varying
neutrino masses. We will show that the two types of models predict
different behaviors of the spatial variation of the neutrino
masses inside Earth and consequently result in different signals
in long baseline neutrino oscillation experiments.

\end{abstract}

\maketitle

\section{Introduction}
There are growing evidences from various cosmic observations,
including type Ia supernova (SNIa) \cite{pelmutter}, cosmic
microwave background (CMB) \cite{spergel}, large scale structures
(LSS) \cite{tegmark,mcdonald}, and so on, that support for a
spatially flat and accelerating universe at the present epoch. In
the context of Friedmann-Robertson-Walker cosmology, this
acceleration is attributed to the so-called dark energy
\cite{turner}. The simplest candidate for the dark energy seems to
be a remnant small cosmological constant. However, many physicists
are attracted to the idea that the dark energy is due to a
dynamical component in the evolution of the universe, such as the
quintessence \cite{wetterich,ratra,frieman,zlatev,steinhardt}, the
K-essence \cite{armendariz,armendariz2,chiba}, the phantom
\cite{caldwell}, or the quintom \cite{feng1,guo,feng2,li2}. \vskip
1ex

Recently there have been a lot of works
\cite{li,li3,gu,fardon,kaplan,bi,gu2,peccei,hli,zhang,guendelman,bi2,brookfield,
horvat2,takahashi,fardon2,blennow,hli2,weiner,honda,bauer,gu3}
which study possible connections between neutrinos and the dark
energy, generally referred to as the neutrino dark energy. One of
the predictions of the class of models of neutrino dark energy is
that neutrino masses are not constant, but can vary as a function
of space and time. This general prediction can be tested with
Short Gamma Ray Burst \cite{hli}, CMB and LSS \cite{brookfield},
and much more interestingly and directly in neutrino oscillation
experiments
\cite{fardon,kaplan,barger3,cirelli,barger4,gonzalez,schwetz}. In
this paper we make a concrete study of the possibility of probing
the property of dark energy and differentiating its dynamic origin
in very long baseline neutrino oscillations. \vskip 1ex

In general for the models of neutrino dark energy, the Lagrangian
is given by
\begin{eqnarray}
\label{L1}
\mathcal{L}= \mathcal{L}^{SM}_{\nu}
+\mathcal{L}_{\phi}+\mathcal{L}_{int},
\end{eqnarray}
where $\mathcal{L}^{SM}_{\nu}$ is the Lagrangian of the standard
model describing the physics of the left-handed neutrinos,
$\mathcal {L}_{\phi}$ is for the dynamical dark energy scalar
$\phi$ such as quintessence or phantom, and $\mathcal {L}_{int}$
describes the sector that mediates the interaction between the
dark energy scalar and neutrinos, and gives rise to variations of
the neutrino masses. \vskip 1ex

At energy much below the electroweak scale, the relevant
Lagrangian for the neutrino dark energy can be written as
\begin{equation}
\label{L2} \mathcal{L}=\mathcal{L}_{\nu} + \mathcal{L}_\phi -c
\sum_{j} m_{j}(\phi)\bar{\nu_{j}}{\nu_{j}},
\end{equation}
where $\mathcal{L}_\nu$ is the kinetic term of neutrinos, $c$ is a
coefficient which takes the value of $1$ for a Dirac neutrino and
$1/2$ for a Majorana neutrino, $m_j(\phi)$ is the scalar field
dependent mass of the $j$-th neutrino that characterizes the
interaction between the neutrino and the dark energy scalar.
\vskip 1ex

The authors of \cite{hli2} have used the recently released SNIa
data to constrain the coupling of the scalar $\phi$ to neutrinos
and the property of the dark energy scalar. They found that the
model with a phantom scalar is mildly favored. However, the data
do not rule out the possibility of the quintessence scalar coupled
to neutrinos. In this paper we will show that these two models
predict different spatial variation of neutrino masses inside
Earth and consequently result in different signals in very long
baseline neutrino oscillations. \vskip 1ex

This paper is organized as follows: in section II we present our
mechanism for the neutrino mass variation; in section III we study
quantitatively the mass-varying effect in the long baseline
neutrino oscillations; Section IV is a brief summary.

\section{Mechanism for variations of neutrino masses }

In the standard model of particle physics, a typical term for the
neutrino masses can be described by a lepton violating dimension-5
operator
\begin{equation}
-\mathcal{L}_{\not L}=\frac{2}{f}l_{L}l_{L}HH+h.c.,
\end{equation}
where $f$ is a scale of new physics beyond the standard model
which generates the $B-L$ violations, $l$ and $H$ are the lepton
and Higgs doublet, respectively. Here we neglect the lepton
generation symbol. When the Higgs field gets a vacuum expectation
value, $\langle H\rangle = v \simeq 174\textrm{GeV}$, the
left-handed neutrino receives a Majorana mass $\sim
\frac{v^{2}}{f}$. In Ref. \cite{gu} the authors proposed a model
where the dark energy scalar $\phi$ couples to the dim-5 operator.
In this model neutrino masses vary along with the evolution of the
universe and the neutrino mass limit imposed by baryogenesis is
modified. \vskip 1ex

The dimension-5 operator above is not renormalizable. It can be
generated from physics beyond the standard model which involves
very heavy particles interacting with standard model particles. At
low energies the heavy particles can be integrated out and thus
resulting in effective, nonrenormalizable terms. For example, in
the model of the minimal see-saw mechanism, we have the neutrino
mass term,
\begin{equation}
-\mathcal{L}=\sum_{ij}h_{ij}\bar{l}_{Li}H\nu_{Rj}+\frac{1}{2}\sum_{ij}M_{ij}\bar{\nu}^{C}_{Ri}\nu_{Rj}+h.c.,
\end{equation}
where $M_{ij}$ are the Majorana mass matrix elements of the
right-handed neutrinos and $h_{ij}$ the Yukawa couplings. The
Dirac masses of the neutrinos are given by $m_{Dij}= h_{ij}v$. Now
integrating out the heavy right-handed neutrinos $\nu_{Rj}$, one
will generate a dim-5 operator as stated above. As pointed out in
Ref. \cite{gu}, there are various possibilities to have the light
neutrino masses varied, such as by coupling the quintessence field
to either the Dirac masses or the Majorana masses of the
right-handed neutrinos, or to both. \vskip 1ex

In this paper we consider the case where the variation of the
neutrino masses is caused by a coupling of the dark energy scalar
$\phi$ to right-handed neutrinos. With the Majorana masses of
right-handed neutrinos varying, $M_{ij}$ becomes a function of the
dark energy scalar field $\phi$, $M_{ij}=M_{ij}(\phi)$.
Furthermore, we assume a linear relationship between the Majorana
mass and $\phi$. Then the relevant Lagrangian can be written as
\begin{equation}
-\mathcal{L}=h_{ij}\bar{l}_{Li}H
\nu_{Rj}+\frac{1}{2}g_{ij}\phi\bar{\nu}^{C}_{Ri}
\nu_{Rj}+h.c.\mp\frac{1}{2} \partial _{\mu}\phi
\partial ^{\mu}\phi +V(\phi),
\end{equation}
where $g_{ij}$ are dimensionless constants and $V(\phi)$ is the
potential for $\phi$. The upper minus sign in $\mp$ is for the
case of quintessence while the lower plus sign for the phantom.
This convention will be used throughout this paper. Now the
Majorana mass matrix elements of the right-handed neutrinos can be
written as
\begin{equation} \label{mr}
M_{ij}=g_{ij}\phi\
\end{equation}
and consequently via the seesaw mechanism we obtain the masses of
the light neutrinos $\nu =\nu_{L}+ \nu_{L}^{C}$,
\begin{equation}
\label{ml1}
m_{\nu} \propto \frac{1}{\phi}.
\end{equation}

Similar to the study on the mass varying neutrinos in
\cite{kaplan}, here we introduce also a coupling between $\phi$
and the baryon matter with the effective potential for $\phi$ at
low energies given by \cite{khoury1,khoury2}
\begin{equation}
\label{V} V^{eff}(\phi)=V(\phi)+\sum_{i}\rho_{i}
e^{\beta_{i}\phi/m_{pl}},
\end{equation}
where $\beta_{i}$ is a dimensionless constant
\cite{damour1,damour2}, $\rho_{i}$ denotes the energy density of
the $i$-th matter field, and $m_{pl}$ is the reduced Planck mass.
The dark energy scalar $\phi$ shall change its value in space
\cite{khoury1,khoury2} according to the equation of motion
\begin{equation}
\label{em}
\bigtriangledown^{2}\phi=\pm\{V_{,\phi}+\sum_{i}\frac{\beta
_{i}}{m_{pl}}\rho_{i}e^{\beta_{i}\phi/m_{pl}}\}.
\end{equation}
In the following we will calculate the evolution of the dark
energy field and the corresponding variation of neutrino masses in
the Earth for both the quintessence and the phantom cases. \vskip
1ex

The density profile of baryon in the Earth is taken as the widely
adopted PREM model \cite{prem}, in which Earth is taken to be
spherically symmetric. The atmosphere is treated as a homogenous
layer of $10~\textrm{km}$ in thickness with a constant density
$\rho_{atm} \simeq 10^{-3}~\textrm{g/cm}^{3}$. Defining $x =
r/R_{\oplus}$ and $R_{atm} = R_{\oplus}+10~\textrm{km}$ with
$R_{\oplus}=6371~\textrm{km}$ the Earth's radius, the baryon
density can be expressed as
\begin{equation}
\label{rho} \rho_{i}(r)=a_{i}+b_{i}x+c_{i}x^{2}+d_{i}x^{3}\;\;\;\;
\textrm{for}\;\;\; r_{i+1}< r\leq r_{i}
\end{equation}
with
\begin{eqnarray}
\label{ai}
(a_{1},a_{2},...,a_{11}) &=& (\rho_{atm},1.02,2.6,2.9,2.691,7.1089,11.2494,5.3197,7.9565,12.5815,13.0885), \\
\label{bi}
(b_{1},b_{2},...,b_{11}) &=& (0,0,0,0,0.6924,-3.8045,-8.0298,-1.4836,-6.4761,-1.2638,0), \\
\label{ci}
(c_{1},c_{2},...,c_{11}) &=& (0,0,0,0,0,0,0,0,5.5283,-3.6426,-8.8381), \\
\label{di} (d_{1},d_{2},...,d_{11}) &=&
(0,0,0,0,0,0,0,0,-3.0807,-5.5281,0)
\end{eqnarray}
in units of $\textrm{g/cm}^{3}$, and
\begin{equation}
\label{ri}
(r_{1},r_{2},...,r_{12})=(R_{atm},R_{\oplus},6368,6356,6346.6,6151,5971,5771,5701,3480,1221.5,0)
\end{equation}
in units of km. We also assume a homogeneous baryon background
outside the atmosphere with the density
\begin{equation}
\label{bu} \rho^{B}_{U}\simeq
4\%\times\rho_{c}\simeq1.4\times10^{-29}\textrm{g/cm}^{3},
\end{equation}
where $\rho_{c} \simeq 4.1 \times 10^{-47}~\textrm{GeV}^4$ is the
critical energy density of the universe at the present epoch.
\vskip 1ex

With these assumptions, Eq. (\ref{em}) can be simplified as
\begin{equation}
\label{em2} \frac{d^{2}\phi}{dr}+\frac{2}{r}\frac{d\phi}{dr}=\pm
\{\frac{dV}{d\phi}+\frac{\lambda _{B}}{m_{pl}}\rho_{B}e^{\lambda
_{B}\phi/m_{pl}}\}
\end{equation}
with the boundary conditions
\begin{equation}\label{boundary11}
\;\;\phi=\phi_{U}\;\;\;\;\textrm{for}\;\;\;  r = r_{c},
\end{equation}
\begin{equation}
\label{boundary12} \frac{d\phi}{dr}=0\;\;\;\;\textrm{for}\;\;\;
r=0.
\end{equation}
Here $r_{c}$ denotes the interface between the static solution and
the cosmological one. For $r > r_{c}$, we expect $\phi$ to become
a constant: $\phi \equiv \phi_{U}$ and $\phi_{U}$ is the value of
the dark energy scalar field on cosmological scales at the present
epoch. For simplicity we take $r_{c}\sim R_{atm}$ which is
consistent with the assumption in Eq. (\ref{bu}) that the baryon
background becomes very thin and homogeneous for $r>R_{atm}$. It
should be noted that we require $\lambda_{B}<10^{-4}$ to satisfy
the equivalence principle constraints \cite{damour1,damour2}.
\vskip 1ex

In addition, we take
\begin{equation}
\label{v} V(\phi)=V_{0}e^{-\beta \phi/m_{pl}}
\end{equation}
as an example \cite{comelli} and then write Eq. (\ref{em2}) as
\begin{equation}
\label{em3}
\frac{d^{2}\phi}{dr^{2}}+\frac{2}{r}\frac{d\phi}{dr}=\pm\{-\frac{\beta}{m_{pl}}V_{0}e^{-\beta
\phi/m_{pl}}+\frac{\lambda _{B}}{m_{pl}}\rho_{B}e^{\lambda _{B}
\phi/m_{pl}}\}.
\end{equation}
For this given potential, the value of $\phi_{U}$ is
\begin{equation}
\label{phiu}
\phi_{U}=\frac{m_{pl}}{\beta}\ln[\frac{2V_{0}}{(1-w)\rho_{\phi}}],
\end{equation}
with $w=\frac{\dot{\phi}^2/2-V}{\dot{\phi}^2/2+V}$ and
$\rho_{\phi}=\frac{1}{2}\dot{\phi}^2+V$ for the quintessence case,
while $w=\frac{-\dot{\phi}^2/2-V}{-\dot{\phi}^2/2+V}$ and
$\rho_{\phi}=-\frac{1}{2}\dot{\phi}^2+V$ for the phantom. Here $w
\simeq -1$ is the equation of the state and $\rho_{\phi} \simeq
73\%\times\rho_{c}\simeq 3.0 \times10^{-47}~\textrm{GeV}^4$ is the
energy density of the dark energy at the present
time\footnote{Here we do not expect the dark energy model to
preserve the tracking behavior \cite{zlatev,steinhardt} and in
this sense, the dark energy scalar field is essentially like the
cosmological constant on the largest scales. This is also
consistent with the static condition adopted in Eq. (\ref{em}).}.
\vskip 1ex

Assuming $\phi/m_{pl}\ll 1$ for $0 \leq r\leq R_{atm}$, which we
will show in the later numerical results is reasonable, we can
simplify Eq. (\ref{em3}) as
\begin{equation}
\label{em4}
\frac{d^{2}\phi}{dr}+\frac{2}{r}\frac{d\phi}{dr}=\pm\{-\frac{\beta}{m_{pl}}\frac{1}{2}(1-w)\rho_{\phi}+\frac{\lambda
_{B}}{m_{pl}}\rho_{B}\}
\end{equation}
and obtain the solution
\begin{equation}
\label{phi} \phi = \left\{
\begin{matrix}
\phi_{U}\;\;\;\;\;\;\;\;\;\;\;\;\;\;\;\;\;\;\;\;\;\;\;\;\;\;\;\;\;\;\;\;\;\;\;\;\;\;\;\;\;\;\;\;\;\;\;\;\;\;\;\;\;\;\;\;\;\;\;\;\;\;\;\;\;\;\;\;\;\;\;\;\;\;\;\;\;\;\;\;\;\;\;\;\;\;\;\;\;\;\;\;\;\;\;\;\;\;\;\;\;\;\;\;\;\;\;\;\;\;\;\textrm{
for}\; r\geq R_{atm} \cr
g_{i}-\frac{f_{i}}{x}\pm\{-\frac{\beta}{m_{pl}}\frac{1}{2}(1-w)\rho_{\phi}R_{\oplus}^{2}\frac{x^2}{6}+\frac{\lambda_{B}}{m_{pl}}R_{\oplus}^{2}(\frac{a_{i}}{6}x^2+\frac{b_{i}}{12}x^3+\frac{c_{i}}{20}x^4+\frac{d_{i}}{30}x^5)\}\;\;\;\;\;\;\;\;\textrm{for}\;
r_{i+1}\leq r\leq r_{i}
\end{matrix}
\right.
\end{equation}
with
\begin{equation}
\label{f} f_{i}= \left\{
\begin{matrix}
\;\;0\qquad\;\;\;\;\;\;\;\;\;\;\;\;\;\;\;\;\;\;\;\;\;\;\;\;\;\;\;\;\;\;\;\;\;\;\;\;\;\;\;\;\;\;\;\;\;\;\;\;\;\;\;\;\;\;\;\;\;\;\;\;\;\;\;\;\;\;\;\;\;\;\;\;\;\;\;\;\;\;\;\;\;\;\;\;\;\;\;\;\;\;\;\;\;\;\;\;\;\;\;\;\;\;\;\;\;\;\;
{\rm for}\;\; i=11 \cr
\;\;f_{i+1}\pm\frac{\lambda_{B}}{m_{pl}}R_{\oplus}^{2}(\frac{a_{i+1}-a_{i}}{3}x_{i+1}^3+\frac{b_{i+1}-b_{i}}{4}x_{i+1}^4+\frac{c_{i+1}-c_{i}}{5}x_{i+1}^5+\frac{d_{i+1}-d_{i}}{6}x_{i+1}^6)\;\;\;\;\;\hspace{0.3ex}\rm{for}\;\;i\leq10
\end{matrix}
\right.
\end{equation}
and
\begin{equation}
\label{g} g_{i}= \left\{
\begin{matrix}
\phi_{U}+\frac{f_{1}}{x_{1}}\pm\{\frac{\beta}{m_{pl}}\frac{1}{2}(1-w)\rho_{\phi}R_{\oplus}^{2}\frac{x_{1}^2}{6}-\frac{\lambda_{B}}{m_{pl}}R_{\oplus}^{2}(\frac{a_{1}}{6}x_{1}^2+\frac{b_{1}}{12}x_{1}^3+\frac{c_{1}}{20}x_{1}^4+\frac{d_{1}}{30}x_{1}^5)\}\;\;\;\;
{\rm for}\;\; i=1 \cr
g_{i-1}\mp\frac{\lambda_{B}}{m_{pl}}R_{\oplus}^{2}(\frac{a_{i+1}-a_{i}}{2}x_{i}^2+\frac{b_{i+1}-b_{i}}{3}x_{i}^3+\frac{c_{i+1}-c_{i}}{4}x_{i}^4+\frac{d_{i+1}-d_{i}}{5}x_{i}^5)\;\;\;\;\;\;\;\;\;\;\;\;\;\;\;\;\;\;\;\;\;\rm{for}\;\;i\geq
2.
\end{matrix}
\right..
\end{equation}
Here we have adopted the definition of $x_{i}$: $x_{i}\equiv
r_{i}/R_{\oplus}$. \vskip 1ex

\begin{figure}
\includegraphics[scale=1.3]{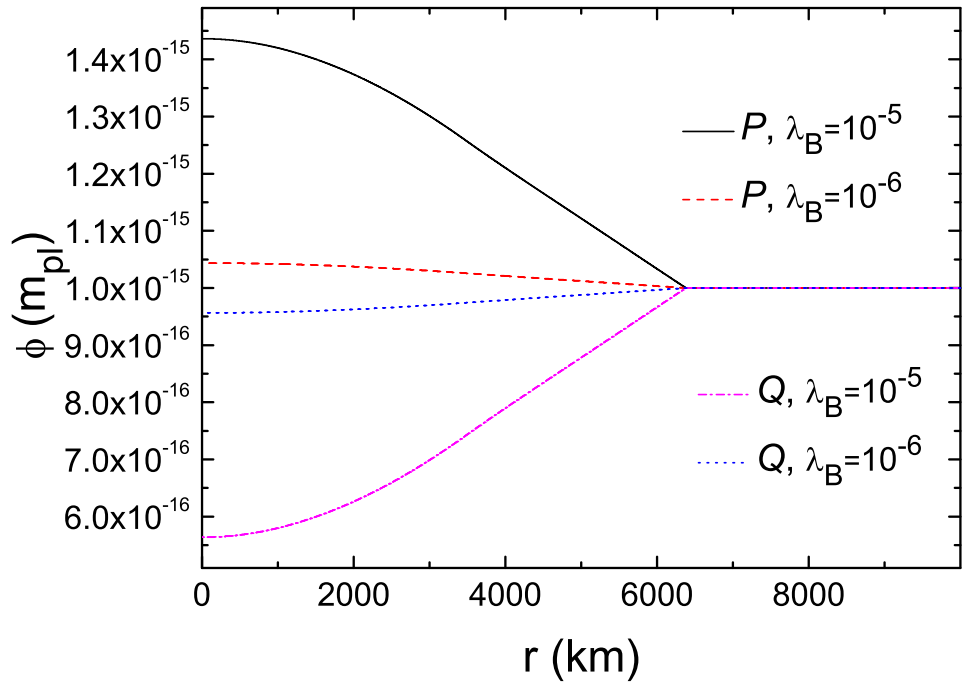}
\caption{\label{phi1}The evolution of the dark energy field $\phi$
in the baryon matter background with $\lambda_{B}=10^{-5}$ and
$10^{-6}$ for the quintessence and the phantom cases,
respectively. It is shown that the assumption of $\phi/m_{pl}\ll
1$ can be fulfilled in the whole space. $P$ and $Q$ are for the
quintessence and phantom cases, respectively.}
\end{figure}

\begin{figure}
\includegraphics[scale=1.3]{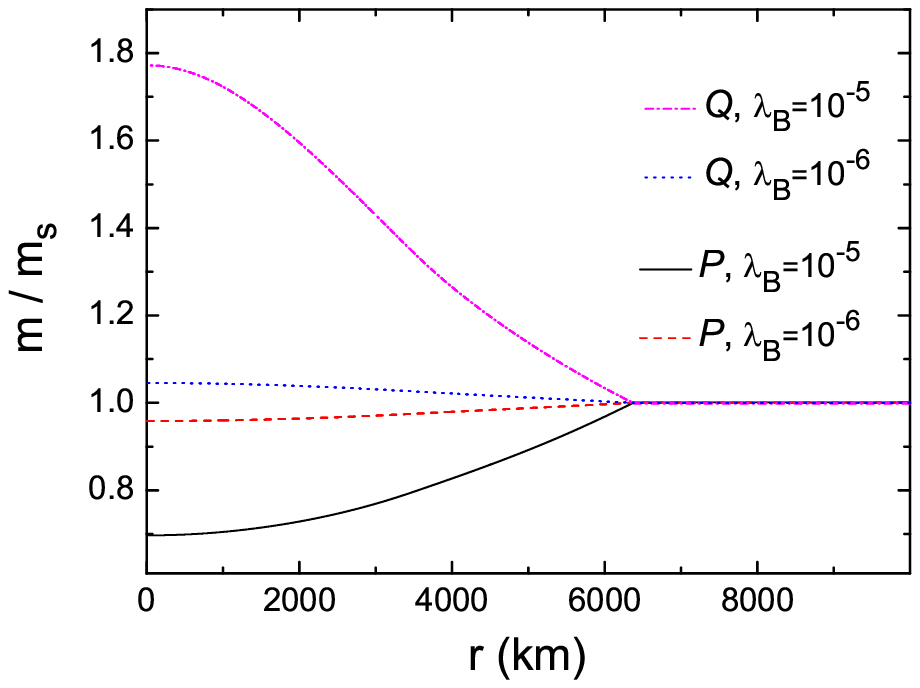}
\caption{\label{mass1}The variation of the neutrino masses in the
baryon background with $\lambda_{B}=10^{-5}$ and $10^{-6}$ for the
quintessence and the phantom cases, respectively. Here $m$ is the
masses of the left-handed Majorana neutrinos and $m_{s}$ denotes
its value on the Earth surface. $P$ and $Q$ are for the
quintessence and phantom cases, respectively.}
\end{figure}

In the numerical calculation, we take $\beta=1$ and
$\phi_{U}\simeq 10^{-15}m_{pl}$ by choosing $V_{0}$ to satisfy the
cosmological observations, $\rho_{\phi}\simeq 3.0
\times10^{-47}~\textrm{GeV}^4$ and $w\simeq -1$. In Fig.
\ref{phi1}, we plot $\phi$ as a function of $r$ with
$\lambda_{B}=10^{-5}$ and $10^{-6}$ for the quintessence and
phantom cases, respectively. Our results show that in the two
cases the variations of the dark energy field can be sizable
inside the Earth. Accordingly, as shown in Fig. \ref{mass1}, the
neutrino masses could have a significant variation inside the
Earth. Meanwhile we notice that the consequences of the
quintessence and the phantom are different due to the opposite
behaviors of the dark energy fields.

\section{Mass-varying effect in long baseline neutrino oscillations}
Now we discuss the mass-varying effect induced by the evolution of
the dark energy in neutrino oscillations. The neutrino propagation
in matter is governed by the Shr$\ddot{o}$dinger equation:
\begin{equation}\label{schr}
i\frac{d}{dt}\left(\begin{array}{c}
\nu_e \\
\nu_\mu \\
\nu_\tau
\end{array} \right)=
\left[[\frac{\phi_{s}}{\phi(x)}]^{2}\ \frac{1}{2E}U
\left(\begin{array}{ccc}
0    &   0                & 0     \\
0    & \Delta m_{21}^{2}  & 0     \\
0    &   0                & \Delta m_{31}^2
\end{array} \right) U^{\dagger} + \sqrt{2}\, G_{F}
\left(\begin{array}{ccc}
N_{e}(x)  &  ~ 0    &~~~ 0     \\
0         &  ~ 0    &~~~ 0     \\
0         &  ~ 0    &~~~ 0
\end{array} \right) \right]
\left(\begin{array}{c}
\nu_{e} \\
\nu_{\mu}\\
\nu_\tau
\end{array} \right).
\end{equation}
Here $U$ is the usual 3-flavour vacuum mixing matrix and
$\sqrt{2}\, G_{F}N_{e}(x)$ is the MSW term
\cite{wolfenstein,mikheyev}. We have also adopted the definition
that $\phi_{s}$, $\Delta m_{ij}^{2}$ are the values of the dark
energy field and the neutrino mass squared differences on the
Earth surface, respectively. In the following numerical estimate,
we will take $\Delta m_{21}^{2}\simeq 7.9\times
10^{-5}~\textrm{eV}^{2}$, $\tan^{2}\theta_{12}\simeq 0.4$ from
KamLAND \cite{kamland}, $\Delta m_{31}^{2}\simeq 2.8\times
10^{-3}~\textrm{eV}^{2}$, $\sin^{2}2\theta_{23}\simeq 1.0$ from
K2K \cite{k2k}, $\sin^{2}2\theta_{13}\le 0.1$ from Chooz
\cite{chooz} which we will take to be zero for simplicity, and a
zero Dirac CP phase in Eq. (\ref{schr}). \vskip 1ex

\begin{figure}
\includegraphics[scale=0.33]{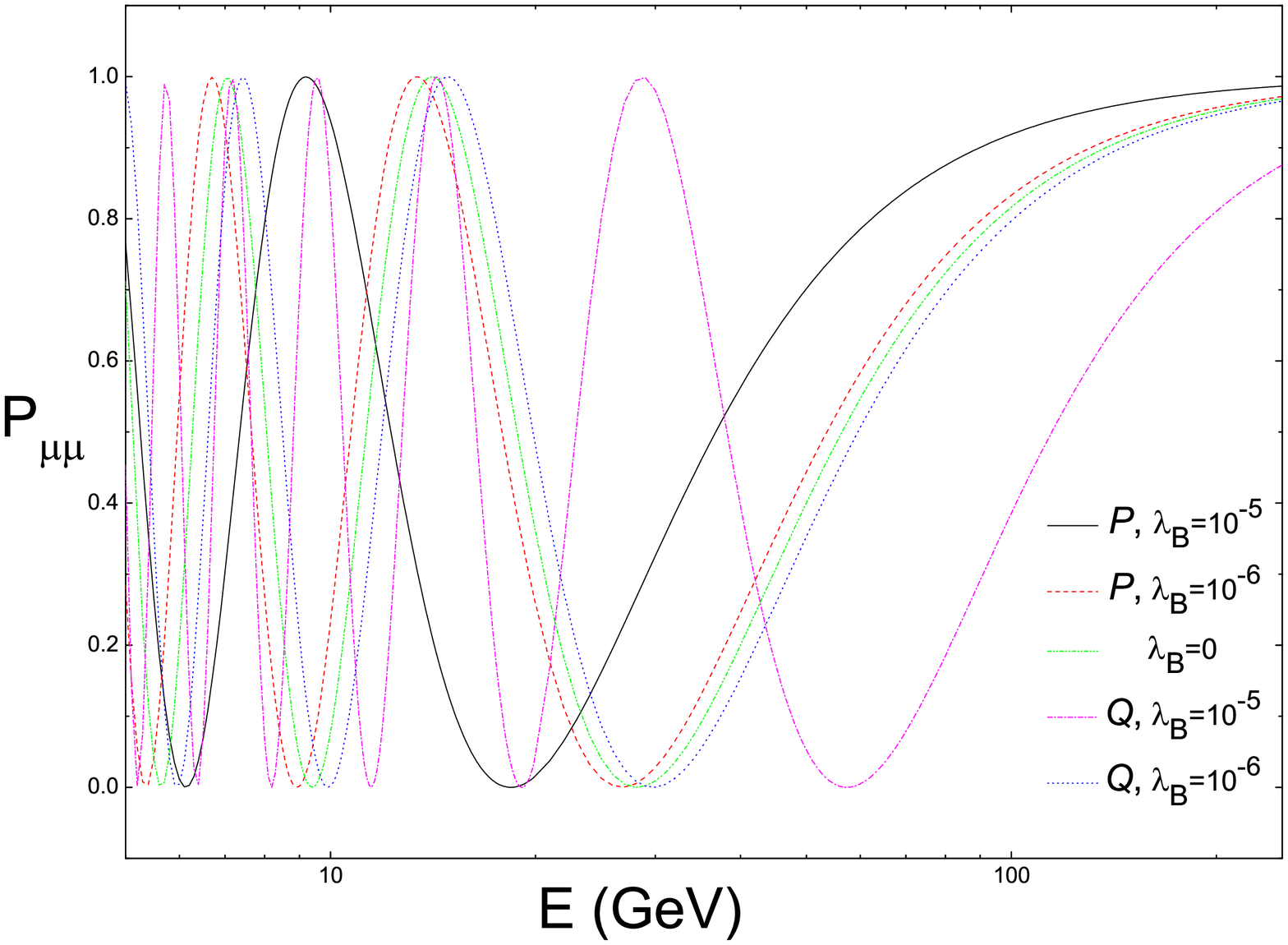}
\includegraphics[scale=0.47]{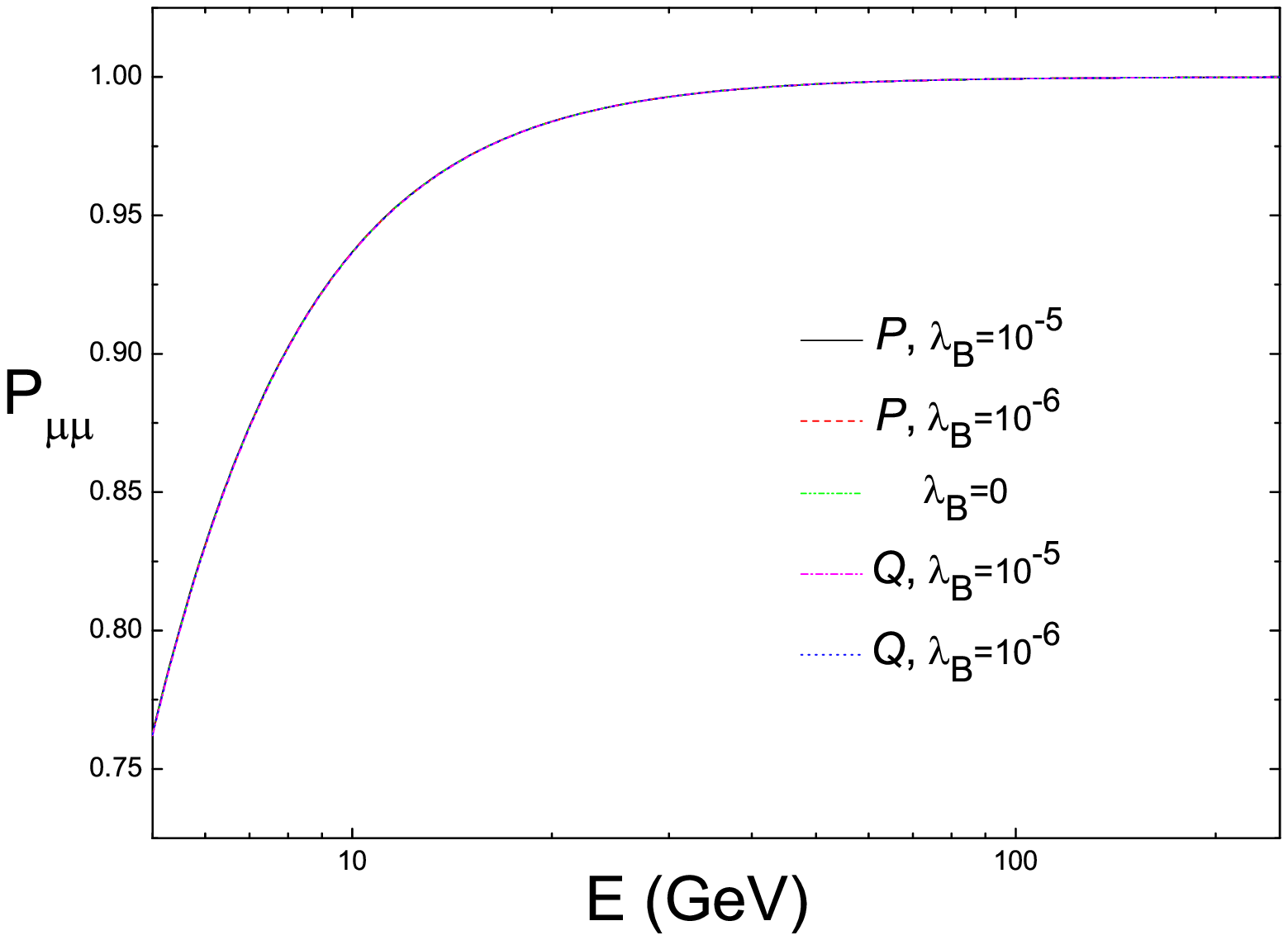}
\caption{\label{prob1}The survival probabilities $P_{\mu\mu}$ of
$\nu_{\mu}$ in the long baseline $L=2R_{\oplus}$ and $L=732~{\rm
km}$ with $\lambda_{B}=10^{-5}$, $10^{-6}$ and $0$. $P$ and $Q$
are for the quintessence and phantom cases, respectively. Note
that $\lambda_\beta = 0$ is identical to the case of decoupling
between the neutrino mass and dark energy. In the calculations, we
have used $\Delta m_{21}^{2}\simeq 7.9\times
10^{-5}~\textrm{eV}^{2}$, $\Delta m_{31}^{2}\simeq 2.8\times
10^{-3}~\textrm{eV}^{2}$, $\tan^{2}\theta_{12}\simeq 0.4$,
$\sin^{2}2\theta_{23}\simeq 1.0$, $\sin^{2}2\theta_{13}\simeq 0$,
and a zero CP phase. The left panel if for $L=2R_\oplus$ and the
right panel $L=732~{\rm km}$}
\end{figure}

\begin{figure}
\includegraphics[scale=0.45]{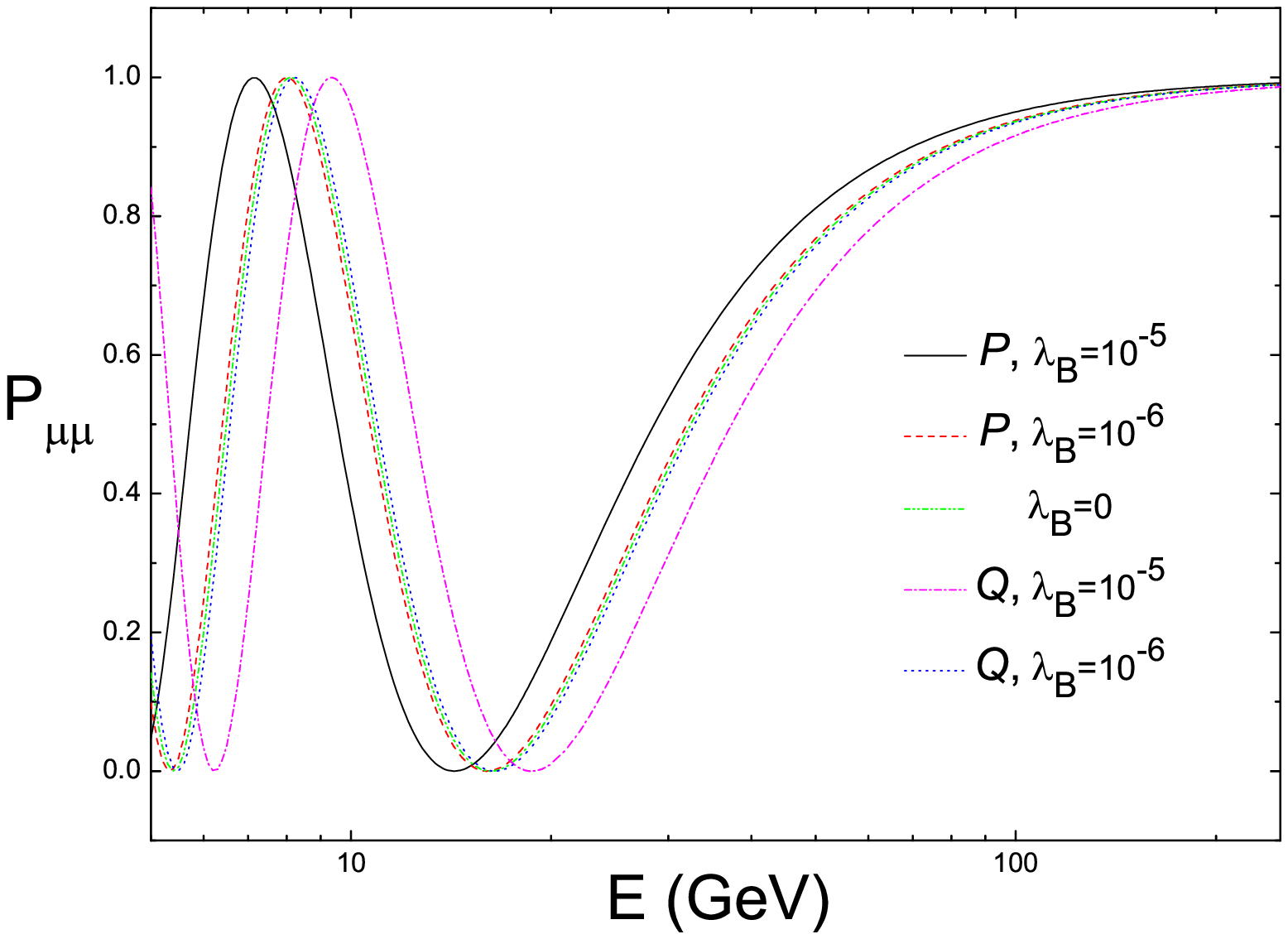}
\includegraphics[scale=0.45]{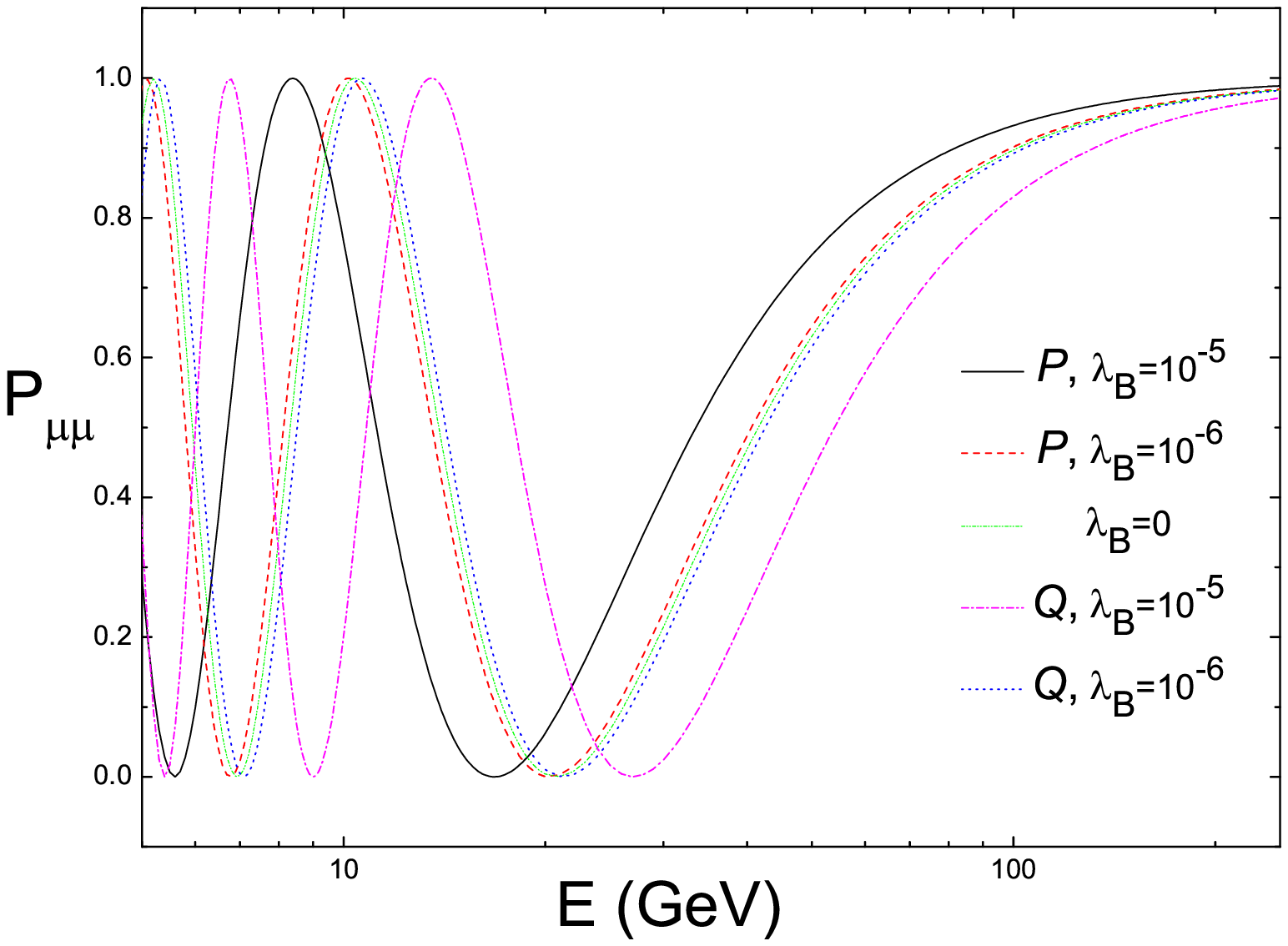}
\caption{\label{prob2}The survival probabilities $P_{\mu\mu}$ of
$\nu_{\mu}$ in the very long baseline $L=7332~\textrm{km}$ and
$L=9400~{\rm km}$ with $\lambda_{B}=10^{-5}$, $10^{-6}$, and $0$.
$P$ and $Q$ are for the quintessence and phantom cases,
respectively. Note that $\lambda_\beta = 0$ is identical to the
case of decoupling between the neutrino mass and dark energy. In
the calculations, we have used $\Delta m_{21}^{2}\simeq 7.9\times
10^{-5}~\textrm{eV}^{2}$, $\Delta m_{31}^{2}\simeq 2.8\times
10^{-3}~\textrm{eV}^{2}$, $\tan^{2}\theta_{12}\simeq 0.4$,
$\sin^{2}2\theta_{23}\simeq 1.0$, $\sin^{2}2\theta_{13}\simeq 0$,
and a zero CP phase. The left panel is for $L=7332~{\rm km}$ and
the right panel for $L=9400~{\rm km}$.}
\end{figure}

In Fig. \ref{prob1}, by taking the longest baseline $L=2
R_{\oplus}$ which appears in the atmospheric neutrino oscillations
and $L=732~\textrm{km}$ which is Fermilab to Soudan or CERN to
Gran Sasso (right panel), we plot the survival probabilities of
$\nu_{\mu}$ with $\lambda_{B}=10^{-5}$, $10^{-6}$ and $0$ for the
case of the quintessence and the phantom, respectively. For
$\lambda_{B}=0$, the spatial variation of the dark energy is
negligible and the neutrino oscillation is identical to the case
of decoupling between the neutrino mass and dark energy. For
$\lambda_{B}=10^{-5}$, it is clear that the survival probabilities
differ significantly for the cases of the quintessence, phantom,
and decoupling of neutrino and dark energy for $L=2R_\oplus$. But
for $L=732~{\rm km}$ the different cases can not be distinguished.
\vskip 1ex

We also consider the very long baseline $L=7332~\textrm{km}$ which
is the distance from Fermilab to Gran Sasso underground laboratory
in Italy \cite{barger5} and $L=9400~{\rm km}$ which is the
distance from Fermilab to Beijing. As shown in Fig. \ref{prob2},
the survival probabilities $P_{\mu\mu}$ with $\lambda_{B}=10^{-5}$
is sensitive enough to distinguish the cases of the quintessence,
phantom, and decoupling. The most sensitive measurement is perhaps
at the second zero of the survival probability. For $L=9400~{\rm
km}$, the decoupling case has the zero at about muon neutrino
energy of 20 GeV, the quintessence above 20 GeV, and the phatom
below 20 GeV. The separation in energy is sufficiently large that
it should make the distinction of the three case clean.

\section{Summary}
In this paper, we discuss a possibility of studying the dark
energy property with long baseline neutrino oscillation
experiments. We consider two types of models of neutrino dark
energy where for one model the scalar is taken to be
quintessence-like and for another model phantom-like. These
scalars couple to the neutrinos which give rise to a variation of
the neutrino masses. We take a specific scalar dark energy
potential with the Earth baryon background and then calculate the
spatial variation of the dark energy field for the case of the
quintessence and the phantom, respectively. We find the
corresponding evolution behaviors of the neutrino masses inside
the Earth could be significantly different in the two cases and
hence the property of the dark energy may be probed in the long
baseline neutrino oscillations.

\newpage
\begin{acknowledgments}
This work is supported in part by the NSF of China under the grant
No. 10575111, 10105004, 10120130794, 19925523, 90303004 and also
by the Ministry of Science and Technology of China under grant No.
NKBRSF G19990754.
\end{acknowledgments}

\end{document}